\documentclass[proceedings, preprint]{rmaa}

\usepackage{paralist}
\usepackage{psfrag,color}

\SetYear{2010}
\SetConfTitle{XIII Latin American Regional IAU Meeting}

\title{Iron Depletion in Ionized Nebulae of the Large Magellanic Cloud \altaffilmark{1}}

\author{
  G. Delgado-Inglada,\altaffilmark{2} 
  M. Rodr\'iguez ,\altaffilmark{2}
  J. Garc\'ia-Rojas, \altaffilmark{3}
  M. Pe\~na,\altaffilmark{4}
  and M. T. Ruiz, \altaffilmark{5}}

\altaffiltext{1}{Based on data obtained with the 6.5-m Magellan-Clay 
telescope at Las Campanas Observatory (Carnegie Institution), Chile.}
\altaffiltext{2}{Instituto Nacional de Astrof\'isica, \'Optica y Electr\'onica (INAOE), 
Apdo Postal 51 y 216, 72000 Puebla, Mexico (gloria, mrodri@inaoep.mx).}
\altaffiltext{3}{Instituto de Astrof\'isica de Canarias (IAC), C/V\'ia L\'actea s/n, E38205 La Laguna, Tenerife, Spain (jogarcia@iac.es).}
\altaffiltext{4}{Instituto de Astronom\'ia, Universidad Nacional Aut\'onoma de M\'exico, Mexico 
(miriam@astroscu.unam.mx).}
\altaffiltext{5}{Departamento de Astronom\'ia, Universidad de Chile, Chile (mtruiz@uchile.cl).}

\shortauthor{Delgado-Inglada, Rodr\'iguez, Garc\'ia-Rojas et al.}
\shorttitle{Iron Depletion in Ionized Nebulae of the LMC}

\listofauthors{G. Delgado-Inglada, M. Rodr\'iguez, J. Garc\'ia-Rojas, 
M. Pe\~na, \& M. T. Ruiz}
\indexauthor{Delgado-Inglada, G.}
\indexauthor{Rodr\'iguez, M.}
\indexauthor{Garc\'ia-Rojas, J.}
\indexauthor{Pe\~na, M.}
\indexauthor{Ruiz, M. T.}

\abstract{We present here preliminary results of an analysis of the iron abundance in the
ionized gas of three planetary nebulae and one \ion{H}{2} region of the Large Magellanic Cloud
(LMC). These results are compared with the ones we obtain for a sample of Galactic and
extragalactic nebulae. We find that the amounts of iron depletion into dust grains in LMC
nebulae are similar to those found in Galactic nebulae. Objects with lower metallicities
show lower depletions, but a larger sample of objects is needed to explore the reasons
behind this trend.}

\resumen{Presentamos resultados preliminares de un an\'alisis de la abundancia de hierro en
el gas ionizado de tres nebulosas planetarias y una regi\'on \ion{H}{2} de la Gran Nube de
Magallanes (GNM). Comparamos estos resultados con los obtenidos para una muestra de
nebulosas Gal\'acticas y  extragal\'acticas. Encontramos que los factores de depleci\'on
del hierro en granos de polvo en las nebulosas de la GNM son similares a los encontrados en
nebulosas de la Galaxia. Los objetos con menor metalicidad muestran depleciones menores,
pero ser\'\i a necesaria una muestra m\'as grande de objetos para explorar las causas de
esta tendencia.}

\addkeyword{dust, extinction}
\addkeyword{galaxies: ISM}
\addkeyword{H~II regions}
\addkeyword{ISM: abundances}
\addkeyword{planetary nebulae: general}

\begin{document}
\maketitle

\section{Introduction}
\label{sec:intro}
Dust is produced in the outflows of evolved stars, and then is injected into the 
interstellar medium, where grains can grow in the densest regions, 
and are eventually destroyed by different mechanisms \citep{Whittet_03}. 
Planetary nebulae (PNe) and \ion{H}{2} regions are powerful tools to study the 
life cycle of dust grains in the ionized medium. On the one hand, the dust present 
in PNe was recently formed in the cool atmospheres of their progenitor stars. 
On the other hand, \ion{H}{2} regions contain 
grains that were located before in the associated molecular clouds, and can be 
considered processed dust. Therefore, the study of the dust present in ionized 
nebulae of different characteristics will provide clues on the processes responsible
for the formation and evolution of the grains and on the role played by metallicity
in the efficiency of these processes.

One way to study dust in ionized nebulae is through the analysis 
of their iron depletion factor, which is defined as the ratio between the expected 
abundance of iron and the one measured in the gas phase. Recently, we 
performed a homogeneous analysis of the iron abundance in a sample of 8 Galactic 
\ion{H}{2} regions and 48 Galactic PNe, and we found that less than 20\% of their iron 
atoms are in the gas (\citealt{DelgadoInglada_09}, Delgado-Inglada \& 
Rodr\'iguez, in prep.). 
Here, we extend this analysis to a sample of extragalactic ionized nebulae, 
and we explore the behavior of iron depletions at different metallicities.

\begin{figure*}[!t]
\begin{center}
\includegraphics[width=11cm]{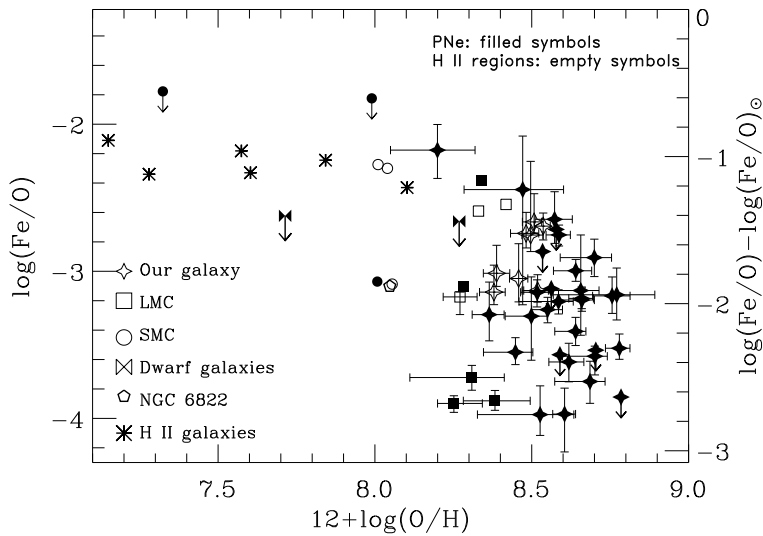}
\vspace{-0.4cm}
\caption{Values of Fe/O (left axis) and the depletion factors 
for Fe/O ($[\mbox{Fe/O}]= \log(\mbox{Fe/O}) - \log(\mbox{Fe/O})_{\odot}$, right 
axis) as a function of the oxygen abundance. Total iron abundances were derived 
using equations 3 and 4 of \citet{Rodriguez_05}. Upper limits represent objects with 
a doubtful identification of the [\ion{Fe}{3}] line 
used in the calculations.}
\label{fig:1}
\end{center}
\end{figure*}

\section{Results}
\label{sec:results}

Deep echelle spectra were obtained for the Large Magellanic Cloud (LMC) PNe SMP~1, SMP~48,
and SMP~85, and the \ion{H}{2} region N11B with the MIKE spectrograph mounted on the 6.5-m
Magellan-Clay  telescope. The wavelength coverage was 3350--5050\AA\ in the blue side, and
4800--9120\AA\ in the red side. Total exposure times on individual objects range from 50 minutes 
to 1 hour and 15 minutes. For each object, we derived a mean electron density using 
between two and four different diagnostic ratios, and two electron temperatures.
We followed the same procedure as for the Galactic objects: we
derived the  total gaseous abundance of iron using the Fe$^{++}$ abundance, and correcting
for  the contribution of higher ionization states with the ionization correction  scheme of
\citet{Rodriguez_05}. We also calculated the iron abundances in a sample of \ion{H}{2} galaxies
and some extragalactic PNe and \ion{H}{2} regions from  the literature
(\citealt{Peimbert_03,Tsamis_03a,Tsamis_03b,Rodriguez_05,
Leisy_06,Kniazev_07,Kniazev_08,Peimbert_05,Izotov_09,LopezSanchez_09}).

Figure~\ref{fig:1} shows our results for
the Fe/O abundance ratio, as well as an estimate of the depletion factor
$[\mbox{Fe/O}]= \log(\mbox{Fe/O}) - \log(\mbox{Fe/O})_{\odot}$, as a function of the
oxygen abundance for all the objects. We refer the reader  to \citet{Rodriguez_05} for a
discussion of the use of the solar abundance ratio $(\mbox{Fe/O})_{\odot}$ to estimate
depletion factors at different metallicities. 

\citet{Rodriguez_05} found a trend of decreasing depletions at lower metallicities.
Figure~\ref{fig:1} shows a similar result but for a larger sample of objects. This trend is
mainly defined by \ion{H}{2} galaxies (see also \citealt{Izotov_06}) and it shows a higher 
dispersion as the metallicity increases. The trend suggests that iron atoms are
significantly less attached to dust grains at low metallicity. This could be due to
changes in the efficiencies of the processes responsible for dust formation, growth or
destruction. 
 
As we mentioned above, PNe can provide information about the efficiency of  circumstellar
dust production. However, data for low metallicity PNe are scarce  and unreliable (since
they are often based on doubtful line identifications). The LMC PNe show iron depletion
factors similar to the ones found in Galactic PNe, but PNe with lower metallicity could
have lower iron  depletions, as in the case of the \ion{H}{2} regions. To answer this
question, and to study the changes with metallicity of the total range of depletion
factors, deep spectra for a larger sample of low-metallicity nebulae, especially PNe, are
needed.


\begin{thebibliography}
\bibitem[Delgado-Inglada et al.(2009)]{DelgadoInglada_09} 
Delgado-Inglada, G., Rodr\'iguez, M., Mampaso, A. \& Viironen, K.\ 2009, \apj, 694, 1335

\bibitem[Izotov et al.(2009)]{Izotov_09} 
Izotov, Y.~I., Guseva, N.~G., Fricke, K.~J., \& Papaderos, P.\ 2009, \aap, 503, 61 

\bibitem[Izotov et  al.(2006)]{Izotov_06}
Izotov, Y.~I., Stasi{\'n}ska, G., Meynet, G., Guseva, N.~G., \& Thuan, T.~X.\ 2006, \aap, 448, 955

\bibitem[Kniazev et al.(2008)]{Kniazev_08} 
Kniazev, A.~Y., et al.\ 2008, \mnras, 388, 1667 

\bibitem[Kniazev et al.(2007)]{Kniazev_07} 
Kniazev, A.~Y., Grebel, E.~K., Pustilnik, S.~A., \& Pramskij, A.~G.\ 2007, \aap, 468, 121 

\bibitem[Leisy \& Dennefeld(2006)]{Leisy_06} 
Leisy, P., \& Dennefeld, M.\ 2006, \aap, 456, 451 

\bibitem[L{\'o}pez-S{\'a}nchez \& Esteban(2009)]{LopezSanchez_09} 
L{\'o}pez-S{\'a}nchez, {\'A}.~R., \& Esteban, C.\ 2009, \apss, 324, 355 

\bibitem[Peimbert(2003)]{Peimbert_03} 
Peimbert, A.\ 2003, \apj, 584, 735 

\bibitem[Peimbert et al.(2005)]{Peimbert_05} 
Peimbert, A., Peimbert, M., \& Ruiz, M.~T.\ 2005, \apj, 634, 1056 

\bibitem[Rodr\'iguez \& Rubin(2005)]{Rodriguez_05} 
Rodr\'iguez, M. \& Rubin, R.~H.\ 2005, \apj, 626, 900 

\bibitem[Tsamis et al.(2003a)]{Tsamis_03a} 
Tsamis, Y.~G., Barlow, M.~J., Liu, X.-W., Danziger, I.~J., \& Storey, P.~J.\ 2003a, \mnras, 338, 687 

\bibitem[Tsamis et al.(2003b)]{Tsamis_03b} 
Tsamis, Y.~G., Barlow, M.~J., Liu, X.-W., Danziger, I.~J., \& Storey, P.~J.\ 2003b, \mnras, 345, 186 

\bibitem[{{Whittet}(2003)}]{Whittet_03}
{Whittet}, D.~C.~B. 2003, Dust in the Galactic environment (2nd ed.; Bristol: IoP)
\end{thebibliography}
\end{document}